# KM3NeT-ORCA: Oscillation Research with Cosmics in the Abyss

**Paschal Coyle on behalf of the KM3NeT Collaboration**

Centre de Physique des Particules de Marseille (CPPM),
162 Avenue de Luminy, 13288 Marseille. France.
E-mail: coyle@cppm.in2p3.fr

**Abstract**. KM3NeT, currently under construction in the abysses of the Mediterranean Sea, is a distributed research infrastructure that will host a $km^3$-scale neutrino telescope (ARCA) for high-energy neutrino astronomy, and a megaton scale detector (ORCA) for neutrino oscillation studies of atmospheric neutrinos. ORCA is optimised for a measurement of the mass hierarchy, providing a sensitivity of $3\sigma$ after 3-4 years. It will also measure the atmospheric mixing parameters $\Delta m^2_{atm}$ and $\theta_{23}$ with a precision comparable to the NOvA and T2K experiments using both the muon neutrino disappearance and tau neutrino appearance channels. It will provide a measurement of the tau neutrino appearance rate with better than 10% precision, a crucial ingredient for tests of unitarity. It will probe the octant of the mixing angle $\theta_{23}$ via matter resonance effects on neutrinos and antineutrinos crossing the core and mantle, which are largely independent on the CP phase. The observation of neutrino oscillations over a wide range of baselines and energies will provide broad sensitivity to new physics such as non-standard neutrino interactions (NSI) and sterile neutrinos.

## 1. Introduction

The search for an understanding of the fundamental behaviour of neutrinos has been at the cutting edge of scientific research for the last sixty years. A variety of experiments with solar, atmospheric, reactor and accelerator neutrinos, spanning energies from the MeV up to tens of GeVs, have provided compelling evidence that the known flavour eigenstates ($\nu_e, \nu_\mu, \nu_\tau$) mix, implying the existence of non-zero neutrino masses; one of the few experimental evidence hinting towards physics beyond the Standard Model.

In the standard 3-neutrino scheme, the PMNS mixing matrix, which relates the neutrino flavour eigenstates to the mass eigenstates ($\nu_1, \nu_2, \nu_3$), can be parameterised in terms of 3 mixing angles $\theta_{12}$, $\theta_{13}$ and $\theta_{23}$, and a CP-violating phase $\delta$. Oscillation experiments are not sensitive to the absolute value of neutrino masses but do provide measurements of the squared-mass splittings $\Delta m_{ij}^2$ (i,j=1,2,3). The values of all these mixing parameters are now extracted from global fits of available data with a reasonable precision. Despite this tremendous progress, many fundamental properties of the neutrino have yet to be determined: the octant of $\theta_{23}$, the absolute masses, whether they are their own anti-particle (i.e. Majorana or Dirac type), the value of CP phase and finally the neutrino mass hierarchy (NMH).

Large volume atmospheric neutrino detectors have played an important role in our experimental determination of the oscillations parameters. Atmospheric neutrinos, produced in cosmic ray interactions with the atmosphere, provide an unprecedented range of energies (MeV to TeV) and baselines (50 km to 12700 km), not accessible to accelerator and reactor experiments. Next generation experiments such as KM3NeT-ORCA [1], PINGU [2], Hyper-Kamiokande [3] are planned to exploit these features with very large volume neutrino detectors over the next decade.

The KM3NeT Collaboration is in the process of building a new research infrastructure consisting of a network of deep-sea neutrino telescopes in the Mediterranean Sea. A phased and distributed implementation is pursued which maximises the access to regional funds, the availability of human resources and the synergetic opportunities for the earth and sea sciences community. The project was recently selected for the 2016 ESFRI roadmap.

The KM3NeT Phase 2 infrastructure will consist of three so-called 'building blocks'. Each building block constitutes a 3-dimensional array of photo sensors that can be used to detect the Cherenkov light produced by relativistic particles emerging from neutrino interactions. Two building blocks will be configured to fully explore the IceCube diffuse signal for cosmic neutrinos with different methodology, improved resolution and complementary field of view, including the Galactic plane. Collectively, these building blocks are referred to as ARCA (*Astroparticle Research with Cosmics in the Abyss*). One building block will be configured to precisely measure atmospheric neutrino oscillations. This building block is referred to as ORCA (*Oscillation Research with Cosmics in the Abyss*). ARCA will be realised at the Capo Passero site in Sicily, Italy and ORCA at the Toulon site in Southern France. Due to KM3NeT's flexible design, the technical implementation of ARCA and ORCA is almost identical. The Letter of Intent for the Phase 2 of KM3NeT describes the physics opportunities and detector design [4].

## 2. ORCA Science

### 2.1. Neutrino Mass Hierarchy

The NMH is termed 'normal' (NH) if $\nu_1<\nu_2<\nu_3$ or 'inverted' (IH) if $\nu_3<\nu_1<\nu_2$, Knowledge of the NMH is an important discriminant between theoretical models of the origin of mass. The NMH serves as an input to cosmological models and neutrino flavour conversion in supernovae explosions. Furthermore, the rates of neutrinoless double beta decay depend strongly on the NMH. Finally, the NMH has a significant impact on the precision determination of the PMNS parameters.

ORCA relies on the presence of matter effects that modify the $\nu_\mu$ survival probability and the rate of $\nu_\mu \Leftrightarrow \nu_e$ appearance at the atmospheric mass scale. The matter effects arise from the $\nu_e$ component of the 'beam' undergoing charged-current elastic scattering interactions with the electrons in the matter. This effectively modifies the observed mixing angles and mass differences in a way that depends on the NMH [5].

Figure 1 shows the expected rate asymmetry, $(N_{IH}-N_{NH})/N_{NH}$, between the NH and IH cases as a function of the energy and cosine of the zenith angle (related to the baseline through the Earth) for both the $\nu_\mu$ and $\nu_e$ events. Detector resolution effects on the reconstructed energy and direction are included. At certain energies and angles the relative flux differences can be as large as 10%; the electron neutrino channel being the most robust against resolution effects.

Figure 2 (Left), shows the expected performance of ORCA to determine the NMH as a function of the assumed $\theta_{23}$ and CP phase. For a true IH the significance is essentially independent of $\theta_{23}$. For a true NH, the significance improves as $\theta_{23}$ increases. If the current value of $\theta_{23}$ from the global fits of around 42° is assumed, ORCA will determine the hierarchy with a median significance of 3 sigma in approximately three years. The ORCA data are relatively insensitive to the CP phase, the significance being reduced by at most 20-30% depending on the true value of $\delta_{CP}$. Figure 2 (Right), shows the median significance as a function of time for a variety of assumptions. As detailed in Ref. [4] many of the possible systematic uncertainties (oscillation parameters, CP phase, overall flux factor, NC scaling, $\nu$/anti-$\nu$ skew, $\mu$/e skew, energy slope) are actually fitted from the data itself when determining the NMH.

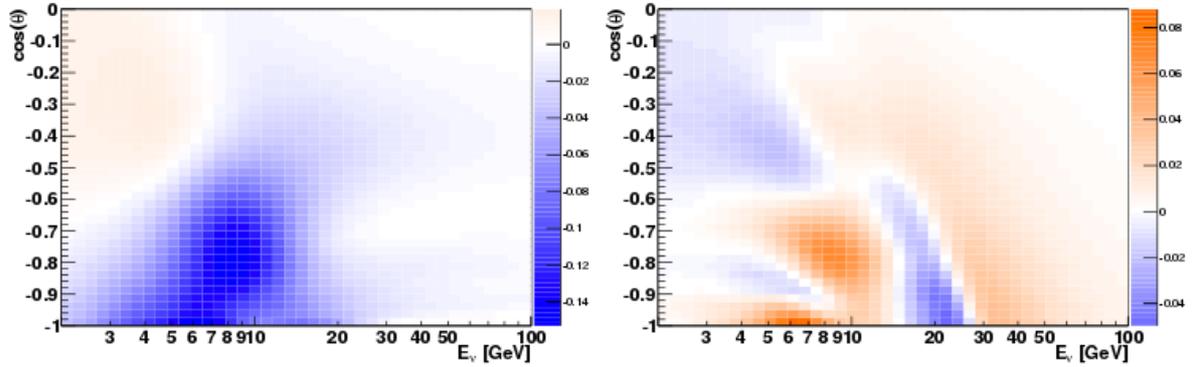

Figure 1: The NMH asymmetry, defined as $(N_{IH}-N_{NH})/N_{NH}$, for $\nu$ and anti-$\nu$ charged current interactions as a function of neutrino energy and cosine zenith angle. Electron neutrinos are on the left and muon neutrinos are on the right. Here the neutrino energy has been smeared by 25% and the zenith angle is smeared by $\sqrt{(M_p/E_\nu)}$.

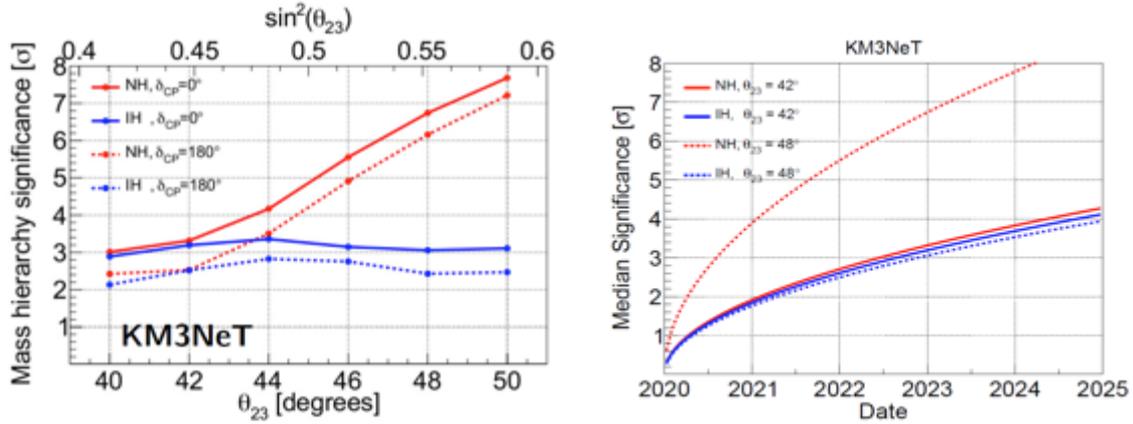

Figure 2: The projected NMH sensitivity for a 115 string ORCA detector. (Left) after 3 years, as a function of $\theta_{23}$. (Right) as a function of time for the indicated scenarios.

*2.2. Oscillation parameters*

Despite providing the first evidence for 'atmospheric' neutrino oscillation, the mixing between the second and third neutrino mass eigenstates is currently the least well measured of the oscillation parameters in the neutrino sector. ORCA will measure $\sin^2\theta_{23}$ and $\Delta m^2_{32}$ parameters via the disappearance of $\nu_\mu$ in the atmospheric flux. Figure 3 shows the expected precision after three years and compares it with current measurements and the expectation from T2K and NOvA at the end of their anticipated data taking in 2020. The precision of ORCA is comparable or better, and is obtained at much higher energies and longer baselines and with very different systematic uncertainties.

The current measurements of $\theta_{23}$ indicate that the angle is close to maximal mixing. If $\theta_{23}$ is not maximal, determining its value and its 'octant' is of importance for understanding the origin of neutrino masses and mixing. Although in two-flavour models, values of $\theta_{23}$ above and below 45° produce identical transition probabilities, this is no longer true for three-flavour oscillation in the presence matter effects. By comparing the rates for neutrinos passing the Earth's core/mantle, ORCA can determine the octant for a wide range of $\theta_{23}$. Unlike the case of T2K and NOvA, the measurement is essentially insensitive to the assumed CP phase.

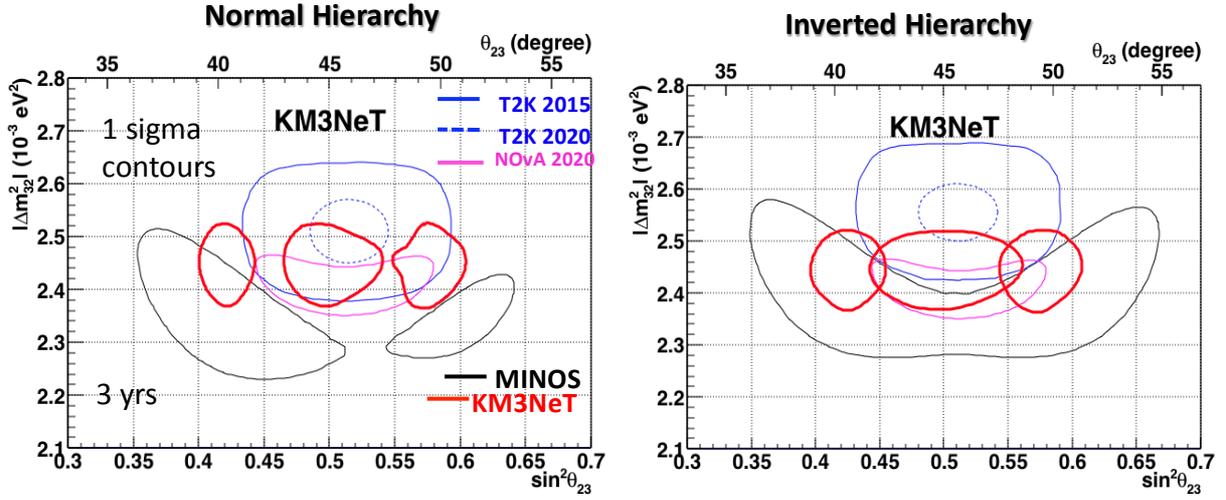

Figure 3: One sigma contours, after three years, of the atmospheric neutrino oscillation parameters for the NH (Left) and IH (Right) scenarios. The current contours for MINOS, T2K and NOvA are indicated, as well as the projected 2020 contours for T2K and NOvA.

*2.3. Tau Appearance*

The unitarity of the PMNS matrix is currently only tested at the 20%-40% level, Many beyond the Standard Model theories, with an extended mixing matrix, could modify the rate of $\nu_\tau$ appearance relative to Standard Model expectations. For vertically upgoing neutrinos, with a baseline of the Earth's diameter, the $\nu_\mu$ disappearance into $\nu_\tau$ is maximum around 24 GeV, well above the ORCA energy threshold. ORCA will detect around 3,000 $\nu_\tau$ CC interactions per year. As shown in Figure 4, this yields a better than 10% precision on the $\nu_\tau$ appearance rate with one year of data taking.

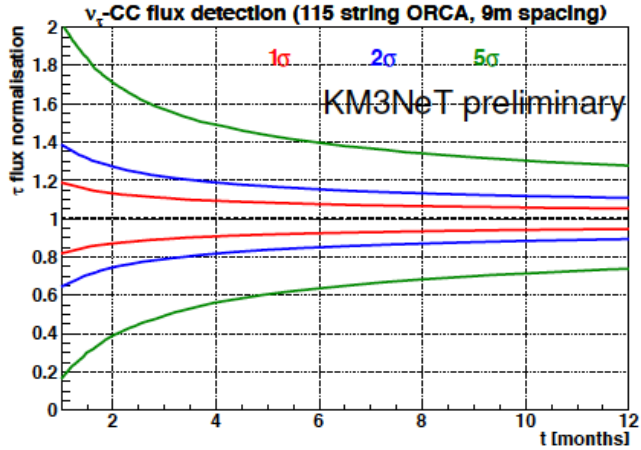

Figure 4: Precision on the rate of $\nu_\tau$ appearance as a function time. The true value is assumed to be the Standard Model expectation (1.0). The corresponding 1, 2, 5 sigma regions are indicated.

*2.4. Sterile neutrinos and non-standard interactions*

Phenomenological extensions of the standard 3ν oscillation framework can include non-standard interactions (NSI) that behave as four-fermion point interactions at low energies. For the sterile neutrino case, the Hamiltonian is extended to four neutrinos and the 4[th] neutrino is assumed not to interact. Figure 5, shows the preliminary estimate of the ORCA sensitivity to sterile neutrinos in the $|U_{\mu 4}|^2$-$|U_{\tau 4}|^2$ parameter space and to NSI in the $\varepsilon_{\tau\tau}$-$\varepsilon_{e\tau}$ parameter space, under some simplified

assumptions [6]. They improve on current limits by an order of magnitude in the NSI parameters and about a factor 5 in the $|U_{\tau 4}|$ mixing.

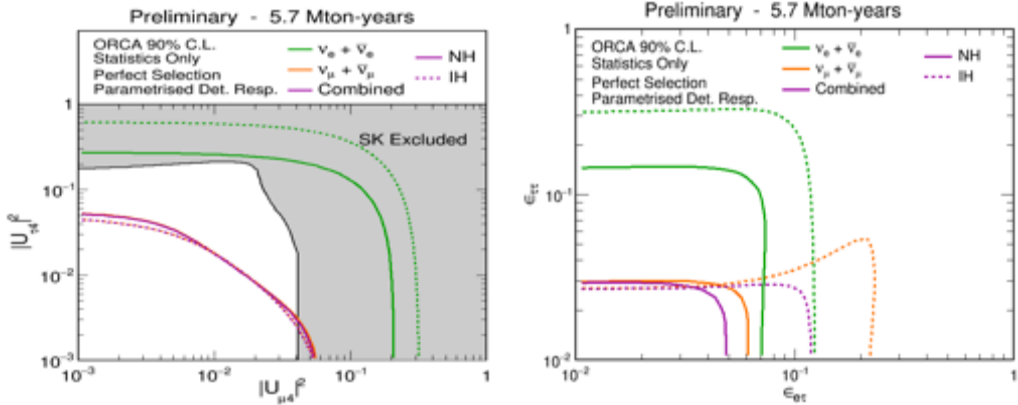

Figure 5: Preliminary one-year sensitivity of ORCA to sterile neutrinos (left) and NSI (Right) in selected slices of the multi-dimensional parameter space. The regions on top-right of the curves are excluded. Limits from Super-Kamiokande are shown for sterile neutrinos.

### 2.5. Other Science: Dark matter, Tomography

Observations in astronomy and cosmology provide irrefutable evidence that the vast majority of the matter in the Universe comprises of non-luminous "dark matter" the nature of which is completely unknown. A theoretically well-motivated candidate is a Weakly Interacting Massive Particle (WIMP). but theory gives little guidance for the mass or whether their interaction with matter is dominantly spin-dependent or spin-independent. WIMPs could be captured in the Sun after scattering off nuclei, accumulate and self-annihilate producing a flux of neutrinos, whose flux and maximum energy depends on the WIMP mass. Since the Sun is primarily made of protons, ORCA can place strong constraints on the spin-dependent WIMP-proton scattering cross-section (Figure 6 (Left)), thereby extending the limits provided by ANTARES, IceCube and Super-Kamiokande to lower WIMP masses.

ORCA will also provide tomographic information on the electron density of the Earth's interior [7]. This new technique is complementary to standard geophysical methods that probe mass density. As shown in Figure 6 (Right) after 10 years of operation, ORCA can measure the electron density in the Earth mantle to an accuracy of ±3.6% (±4.6%) at 1σ confidence level, assuming normal (inverted) mass hierarchy. Here only statistical errors are quoted.

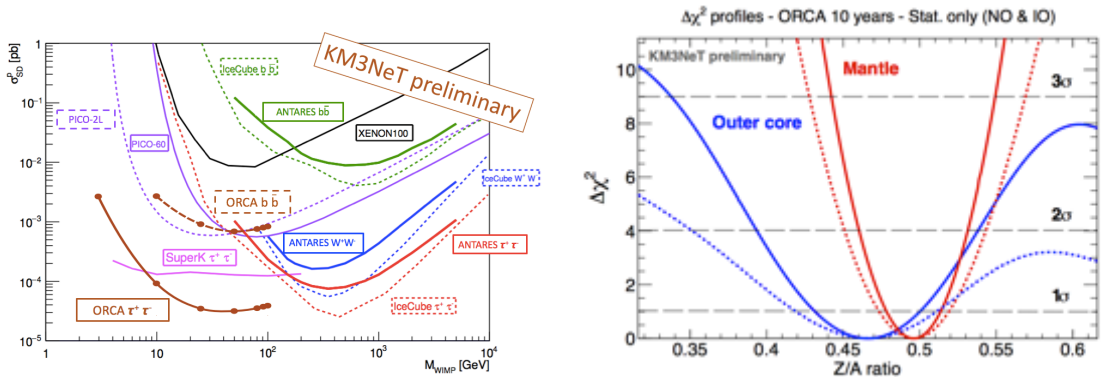

Figure 6: (Left) 90% C.L. limits on the spin-dependent WIMP-proton cross-section after 3 years of data taking, based on counting both $\nu_\mu$ and $\nu_e$ events coming from the direction of the Sun. (Right) Ten year $\Delta\chi^2$ profiles for mantle and outer core for NH (Solid lines) and IH (Dotted lines).

## 3. Detector Design and Technology

To maximise the sensitivity to oscillation phenomena the detector should be capable to distinguish track-like topologies ($\nu_\mu$-CC) from shower-like topologies ($\nu_e$-CC, $\nu_\tau$-CC and neutral current), provide adequate energy and angular resolutions, should be capable to reject the background from downgoing atmospheric muons and large enough to acquire sufficient statistics.

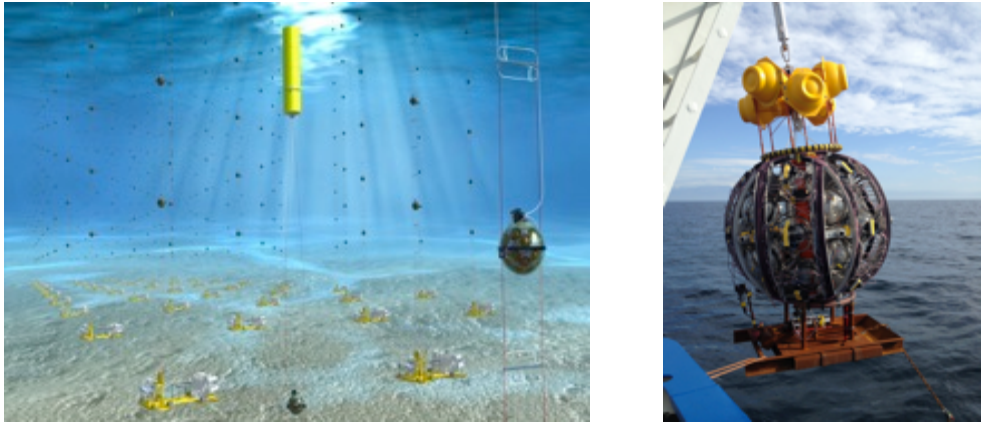

Figure 7: (Left) Artist's impression of a KM3NeT building block. (Right) The spherical frame upon which the detection strings are furled for deployment.

The ORCA infrastructure is located at 42°48'N 06°02'E at a depth of 2450 m, about 40 km offshore from Toulon, France and about 10 km west of the site of the operating ANTARES neutrino telescope [8]. The ORCA array (Figure 7 (Left)) comprises 115 detection strings and each string comprises 18 digital optical modules (DOMs). The vertical spacing is 9 m and the strings are located about 20 m from each other on the seafloor. The total instrumented volume is about 6 Mton. The estimated cost of the ORCA array is about 45 M€ and it is expected to take three years to construct.

The DOMs [9] are distributed in space along flexible strings, one end of which is anchored to the sea floor and the other end is held close to vertical by a submerged buoy. The string comprises two thin parallel ropes that hold the optical modules in place. Attached to the ropes is the vertical electro-optical cable, an oil filled plastic tube which contain the electrical wires and optical fibres used for the power and data transmission. The strings are initially coiled around a spherical frame (Figure 7 (Right)) and deployed by a surface vessel. A remotely operated submersible is used to deploy and connect interlink cables from the base of a string to a junction box. An acoustic signal from the boat triggers an autonomous unfurling of the string from the sea bottom.

A DOM (Figure 8) is a pressure resistant, 17-inch glass sphere containing a total of 31 3-inch PMTs and their associated electronics. The design offers a number of improvements compared to previous designs based on a single large area PMT, most notably: larger photocathode area, digital photon counting, directional information, wider field of view and reduced ageing effects. A position calibration device (acoustic piezo sensor) and a time calibration device (nano-beacon) are also housed inside each sphere. The readout electronics features low power consumption (7 W), high-bandwidth (Gb/s) data transmission using dense wavelength division multiplexing (DWDM), time over threshold measurement of each PMT signal and precision time synchronisation via a white rabbit protocol.

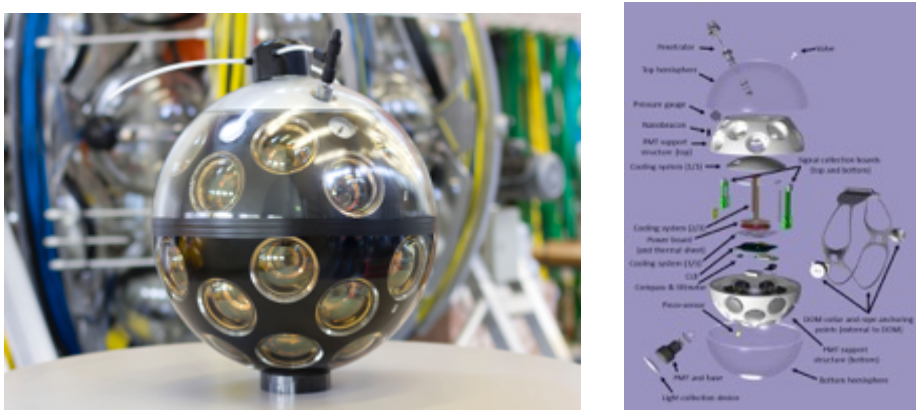

Figure 8 : (Left) photo of a KM3NeT DOM. (Right) exploded view of a DOM interior.

Simulation studies indicate that the ORCA array provides an effective detector mass of about 6 Mton for $\nu_e$ charged current interactions; being fully efficient above 10 GeV and 50% efficient at 4 GeV. It will provide data samples of about 50,000 reconstructed upgoing neutrinos per year for energies below 100 GeV. Reconstruction algorithms applied to $\nu_\mu$-CC and $\nu_e$-CC events yield a Gaussian energy resolution of better than 30% in the range 5-10 GeV. The median angular resolutions on the zenith angle are better than 8° above 5 GeV for both the muon and electron channels; being limited by the kinematic smearing. Discrimination between track and shower topologies is 90% (70%) at 10 GeV for $\nu_e$-CC ($\nu_\mu$-CC). A downgoing muon contamination at the few % level is achieved, while retaining an efficiency of about 80% for the signal.

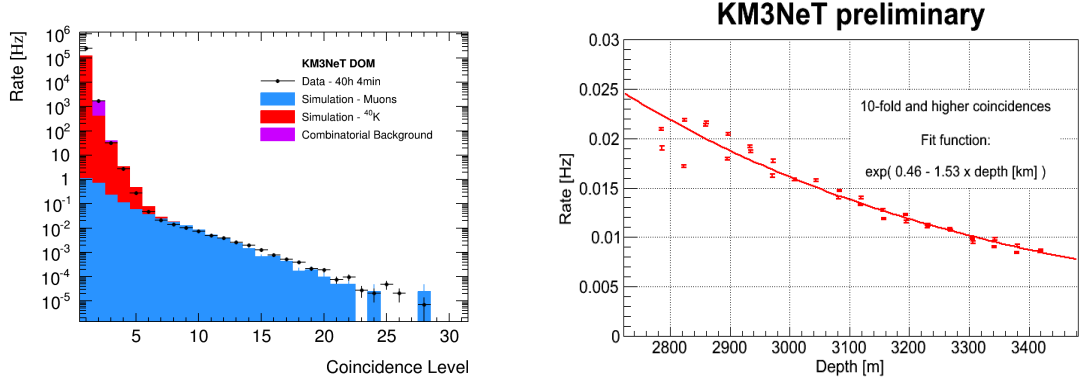

Figure 9: (Left) The rate of events as a function of the coincidence level (number of PMTs with signal in a 20 ns time window). Black dots correspond to data while coloured histograms represent simulations (muons in blue, $^{40}$K in red and accidental coincidences in purple). (Right) The rate of 10-fold coincidences as a function of the DOM depth for two strings.

## 4. Prototype results

The development plan of KM3NeT included the test of a single DOM hosted on an ANTARES line in 2013 [10] and the deployment of a short prototype string with three DOMs in 2014 [11]. In 2016, three full-length ARCA style strings (vertical spacing 36 m) have been deployed at the KM3NeT-Italy site in Capo Passero. Two of these strings are providing high quality data, while the third developed a power issue just after deployment and will be recovered for investigation.

Figure 9 (Left) shows the observed in-situ counting rate versus the multiplicity of PMTs of a single DOM having a hit within a 20 ns time window. At low multiplicities, the counting rate is dominated by light from $^{40}$K decays from the salt in the seawater. The $^{40}$K coincidences provide a powerful method to calibrate the time offsets of the PMTs within a DOM and extract the absolute DOM efficiency, as well as continuously track its dependence as a function of time [12]. At high PMT multiplicities only the Cherenkov light from downgoing muons remain. Figure 9 (Right), shows the rate of 10-fold coincidences as a function of the height of the DOM on the string; the muon attenuation as the depth increases is clearly visible.

## 5. Conclusion

The ORCA array, an instrumented volume of about 5.7 Mton, is optimised for the detection of atmospheric neutrinos in the energy range 3-30 GeV. Its higher energy range, with stronger matter effects and different systematic uncertainties provides complementarity to long-baseline beam or reactor neutrino experiments.

Physics studies demonstrate that the neutrino mass ordering can be determined with a significance of 3-7σ (depending on the true value of the hierarchy and the value of mixing angle $\sin\theta_{23}$) after three years of operation, i.e. as early as 2023. Simultaneously, ORCA will make world-leading measurements of neutrino oscillation parameters, test the maximal mixing hypothesis and provide considerably improved constraints on the unitarity of the PMNS matrix. It will provide excellent sensitivity to potential new physics effects, extend the ANTARES reach for dark matter annihilation to lower energies and establish first tomographic constraints on the composition of the Earth's interior.